# Selection of Optimal Wavelengths for Optical Soiling Modelling and Detection in Photovoltaic Modules

*Leonardo Micheli, Eduardo F. Fernandez, Matthew Muller, Greg P. Smestad, Florencia Almonacid*

## Abstract

Soiling impacts the photovoltaic (PV) module performance by reducing the amount of light reaching the photovoltaic cells and by changing their external spectral response. Currently, the soiling monitoring market is moving toward optical sensors that measure transmittance or reflectance, rather than directly measuring the impact of soiling on the performance of photovoltaic modules. These sensors, which use a single optical measurement, are not able to correct the soiling losses that depend on the solar irradiance spectra and on the spectral response of the monitored PV material. This work investigates methods that can improve the optical detection of soiling, by extracting the full soiling spectrum profiles using only two or three monochromatic measurements. Spectral transmittance data, measured with a spectrophotometer and collected during a 46-week experimental soiling study carried out in Jaén, Spain, was analysed in this work. The use of a spectral profile for the hemispherical transmittance of soiled PV glass is found to significantly improve the soiling detection, returning the lowest errors independently of the PV materials and irradiance conditions. In addition, this work shows that it is also possible to select the measurement wavelengths to minimize the soiling loss detection error depending on the monitored PV semiconductor material (silicon, CdTe, a-Si, CIGS and a representative perovskite). The approaches discussed in this work are also found to be more robust to potential measurement errors compared to single wavelength measurement techniques.

## Graphical Abstract

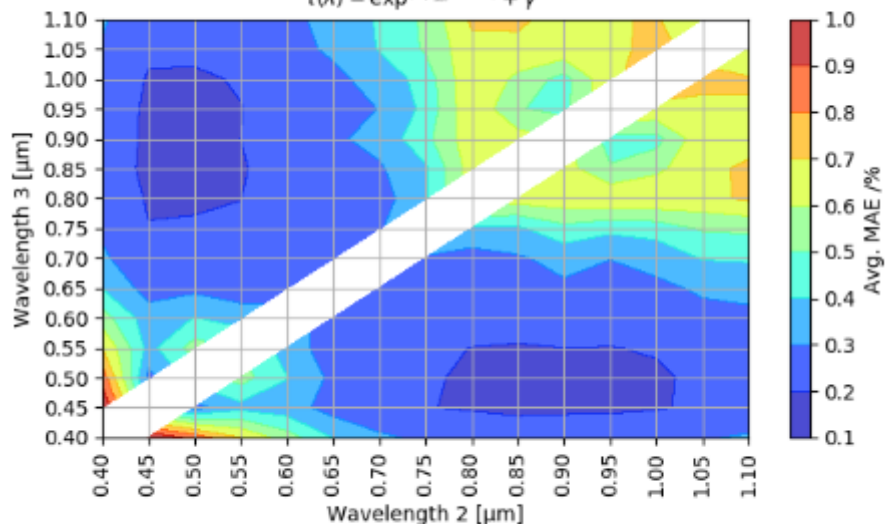

The best results are found for the combination: 0.35 μm, 0.50 μm, and 0.85 μm. The plot above shows the average mean absolute errors found fixing the first measurement (Wavelength 1) at 0.35 μm.

## Nomenclature

*Symbols*

| | |
|---|---|
| $E_G(\lambda, t)$ | spectral solar irradiance in the plane of the PV modules [W/m²/µm] |
| $I_{sc}(t)$ | short-circuit current of the photovoltaic module at time *t* |
| $r_s(t)$ | soiling ratio at a given time *t* |
| $SR(\lambda)$ | spectral response of PV device at wavelength λ |

*Abbreviations*

| | |
|---|---|
| APE | Average Photon Energy [eV] |
| MAE | Mean Absolute Error /% |
| ME | Mean Error /% |
| WST | Waveband Specific Transmittance |
| 2v1e | Two-variable single exponential |
| 3v1e | Three-variable single exponential |

*Greek letters*

| | |
|---|---|
| $\alpha^*$ | wavelength dependence variable in the modified Ångström turbidity equation |
| $\beta^*_{sur}$ | wavelength independent variable in the modified Ångström turbidity equation |
| $\gamma^*$ | offset correction parameter |
| $\lambda$ | Wavelength [µm] |
| $\tau(\lambda, t)$ | relative hemispherical transmittance at wavelength λ at time *t* |
| $\tau_b(t)$ | broadband (average) relative hemispherical transmittance at time *t* |

*Subscripts*

| | |
|---|---|
| meas | measured data point |
| mod | modelled data point |
| ref | reference (clean) module or coupon |
| soil | soiled module or glass coupon |
| *i* | a counting index |

## 1. Introduction

The accumulation of dust, particles, and contaminants on the surface of photovoltaic (PV) modules, a process known as soiling, produces significant losses worldwide [1]. The soiling layer deposited on modules causes forward and backward scattering of some of the sunlight, and increases the portion of light reflected and absorbed, reducing the amount of energy that reaches the photovoltaic cell that can be converted into electricity [2].

The amount of energy produced by a PV module depends on the spectral profile of the incoming solar irradiance that reaches the PV cell and the internal spectral response of the PV cell itself. In addition, the impact of soiling on the transmittance of the PV cover glass is not uniform across the spectrum of sunlight, but varies with the wavelength. More losses are registered in the blue region of the spectrum at shorter wavelengths [3–5]. This means that soiling has a double impact on the incoming solar irradiance: (i) it reduces the broadband intensity and, at the same time, (ii) changes the spectral distribution of the irradiance reaching the PV material [6].

The most common soiling mitigation strategy nowadays is the cleaning of the PV modules [1]. In order to reduce the costs and maximize the profits, a cleaning is generally operated when its cost is lower than the additional revenues gained by the energy recovered [7]. An optimal mitigation strategy therefore requires continual monitoring of the soiling accumulated on the PV modules. Today, soiling is generally monitored through the installation of a soiling station or an optical soiling detector. In the first case, the energy output of a soiled PV device is compared with the energy output of a similar regularly cleaned PV device [8]. This approach allows for the direct measurement of the impact of soiling on the performance of PV modules, but requires periodic maintenance [9]. In order to lower the cost of soiling monitoring, a new class of maintenance-free optical soiling detector products have been launched [10–12]. These devices do not require periodic cleanings and are based on a single optical measurement, which is then converted into an estimated electrical performance loss for the PV module.

The aim of this work, which builds on our previously published studies [2,3], is to analyse the correlations between transmittance loss and electrical losses due to soiling, based on the characteristics of the solar irradiance spectrum and different PV semiconductor materials. These include: monocrystalline silicon (m-Si); poly-crystalline (or multi-crystalline) silicon (p-Si); amorphous silicon (a-Si); Cadmium Telluride (CdTe); Copper Indium Gallium (di)Selenide (CIGS) and a representative perovskite. In the first part, two methods previously presented to estimate the soiling transmittance profiles are employed to model the weekly transmittance losses measured for a 46-week outdoor data collection study in southern Spain. These methods are compared with those based on a single measurement (i.e. monochromatic or average transmittance), which neglect the spectral profile of soiling.

The investigated spectral models require only two or three variables as input, and therefore enable modelling the full soiling transmittance spectra and the associated PV specific electrical loss. For this reason, in the second part of the paper, these models are evaluated for their abilities to estimate the soiling induced electrical loss under different irradiance conditions and for different photovoltaic materials. We compare the results with those obtained when the assumption that the broadband (or average) transmittance or the transmittance at a single wavelength can provide robust estimation of the soiling losses is made. The ultimate goal is to investigate whether it is possible to improve the current optical soiling detection techniques by using two or three single-value transmittance measurements and to determine if this approach makes it possible to estimate the soiling ratio with higher accuracy for different PV technologies under various solar irradiance conditions.

## 2. Background & Motivation

The impact of soiling on the electrical performance of PV modules is generally quantified through the soiling ratio [13], $r_s$, which, at a given time t, is defined as:

$$r_s(t) = \frac{\text{Isc}_{soil}(t)}{\text{Isc}_{ref}(t)} \quad (1)$$

where $\text{Isc}_{soil}(t)$ and $\text{Isc}_{ref}(t)$ are the short-circuit current of the module in natural soiling conditions and the short-circuit current that the same module would generate if no soiling was accumulated on its surface. According to this definition, the soiling ratio changes over time, has a value of 1 for no soiling and decreases as soiling increases. The soiling losses are given by $1 - r_s(t)$. The use of the short-circuit current ratio of Eq. (1), instead of the ratio of the output electrical powers, is possible because of the assumption of uniform soiling [8].

As mentioned, the impact of soiling on the electrical performance of a PV module not only depends on the transmittance of soiling, but also on the spectral response of the PV materials and on the spectral distribution of the irradiance. The same methodology described in the literature [14] and employed in our previous work [3] has been used to calculate the short circuit currents and leads to the following equation:

$$r_s(t) = \frac{\text{Isc}_{soil}(t)}{\text{Isc}_{ref}(t)} = \frac{\int_{\lambda_1}^{\lambda_2} E_G(\lambda,t) \cdot \tau(\lambda,t) \cdot SR(\lambda) \cdot d\lambda}{\int_{\lambda_1}^{\lambda_2} E_G(\lambda,t) \cdot SR(\lambda) \cdot d\lambda} \quad (2)$$

where $\lambda_1$ and $\lambda_2$ are the lower and upper limits, respectively, of the absorption band of the PV material, $E_G(\lambda, t)$ is the spectral distribution of the solar irradiance in the plane of the PV modules, and $\tau(\lambda, t)$ is the hemispherical transmittance due to soiling [2,3]. *SR(λ)* is the spectral response of the photovoltaic material (i.e. Si, CdTe, a-Si).

For this study, we shall neglect the reflection and absorption losses of the glass itself. A more precise analysis would include the transmission of the clean glass inside the integral of both the numerator and denominator of Eq. (2). In the absence of other optical losses, the product, $\tau(\lambda, t) \cdot SR(\lambda)$ is the ideal *external* spectral response of a soiled PV module. The transmittance loss at each wavelength is $1 - \tau(\lambda, t)$. If a constant soiling transmittance profile is assumed, of value $\tau_x(t)$, the soiling ratio takes on the same value as the transmittance, $r_s(t) = \tau_x(t)$. The broadband (average) transmittance at a particular time is $\tau_b(t)$, which for convenience will be shorted to $\tau_b$.

Optical sensors take a single measurement and thus neglect the wavelength dependent effects of soiling, together with the different spectral responses possible for the various types of PV absorber materials. In reality, the spectral profile of the transmittance due to soiling changes depending on the amount and the type of soiling deposited, which varies from site to site and can also vary with time at the same site [15]. In general, the transmission of light due to soiling has a gradually increasing spectral profile [3–5], with larger losses in the blue region, as is shown in Fig. 1. In our previous study [3], we demonstrated that a transmittance measurement at a single wavelength or the use of an average transmittance can be successfully used to identify the soiling profile trends over time, with extreme accuracy ($R^2>0.99$). Each PV absorber material (PV technology) had a preferred wavelength at which the measurement returned the lowest error. For example, the transmittance at 0.6 μm was found to be optimal for m-Si and p-Si. Those measurements are able to rank the severity of soiling and to differentiate between high and low soiling conditions. Despite that, the actual soiling loss calculation can still be subjected to a bias if the spectral transmittance profile of soiling was not flat (see Fig. 1), which can have varied effects on different PV materials and can change under specific irradiance conditions. Because of its shape, taking the spectral profile into account can significantly improve the soiling detection and reduce the error, making it possible to adjust the estimation of the electrical impact of soiling according to the PV material and the irradiance conditions.

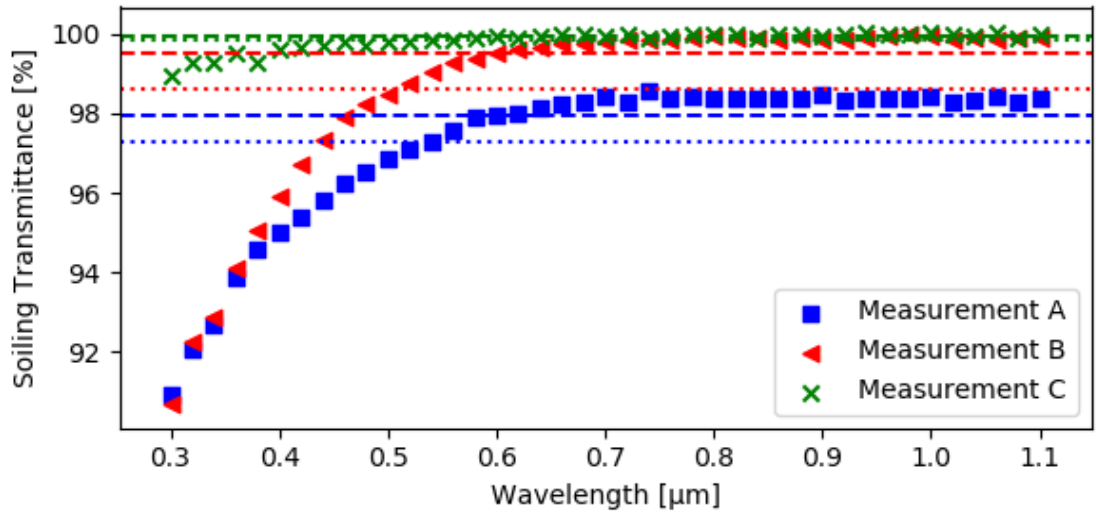

Fig. 1. Examples of the relative hemispherical transmittance spectra of a soiled PV glass coupon (versus clean glass) measured on three different days in Jaén, Spain [3]. The average value of the transmittance (dotted lines), $\tau_b(t)$, and the transmittance values at 0.6 µm (dashed lines) are also plotted as horizontal lines.

PV modules can be made of different PV absorber materials, each having different spectral behaviours. The spectral behaviour of a PV module is established, in part, by its bandgap and is described by its spectral response, which expresses the ratio between the current produced by the PV absorber material and the incident power density at a given wavelength. This means that the same amount of soiling deposited on the PV module's cover glass, with the same transmittance profile, can lead to different losses for each PV material [3–5]. Additionally, sunlight does not have a constant irradiance spectrum. Instead, it is made up of photons of different energies with various intensities. Also, the atmosphere, depending on the climatic conditions and the position of the sun in the sky, can selectively absorb some photons and thus affect the spectral distribution of the irradiance reaching the PV modules. Therefore, to estimate the PV power production with the maximum accuracy, it is essential to take into account both the irradiance hitting the PV material and the PV material's spectral response. Due to the spectral characteristics of its transmittance profile, soiling affects the solar irradiance and changes the spectral distribution reaching the PV solar cell encapsulated in the module. Therefore, the same amount of soiling can produce different losses, even for the same PV absorber materials, depending on the input spectral irradiance.

An example is shown in Fig. 2, where the effect of the same soiling transmittance shown as Measurement A in Fig. 1 is modelled using Eqs. (1) and (2) for two silicon-based PV materials under two different irradiance conditions. These materials are m-Si and a-Si. Prior work [3] established that the optimum single-value wavelength for the measurement of soiling for m-Si devices is 0.6 µm. For the relative hemispherical transmittance indicated as A, the transmittance loss at 0.6 µm is 2.0%. Measurement A yields an average transmittance loss of 2.7%, but, from the soiling ratio of Eq. (2), can cause electrical losses between 1.9% and 2.9%, depending on the irradiance and the PV material (see Fig. 2). This means that, in this case, assuming an electrical output power loss for a soiled PV module equal to a transmittance loss can result in a large relative error. It is important to mention that this error is expected to grow as the broadband (average) transmittance loss increases. This expectation is due to the fact that, as shown in previous works [3–5,16], the difference between losses at short and long wavelengths tends to increase with the severity of soiling (from low to high soiling conditions).

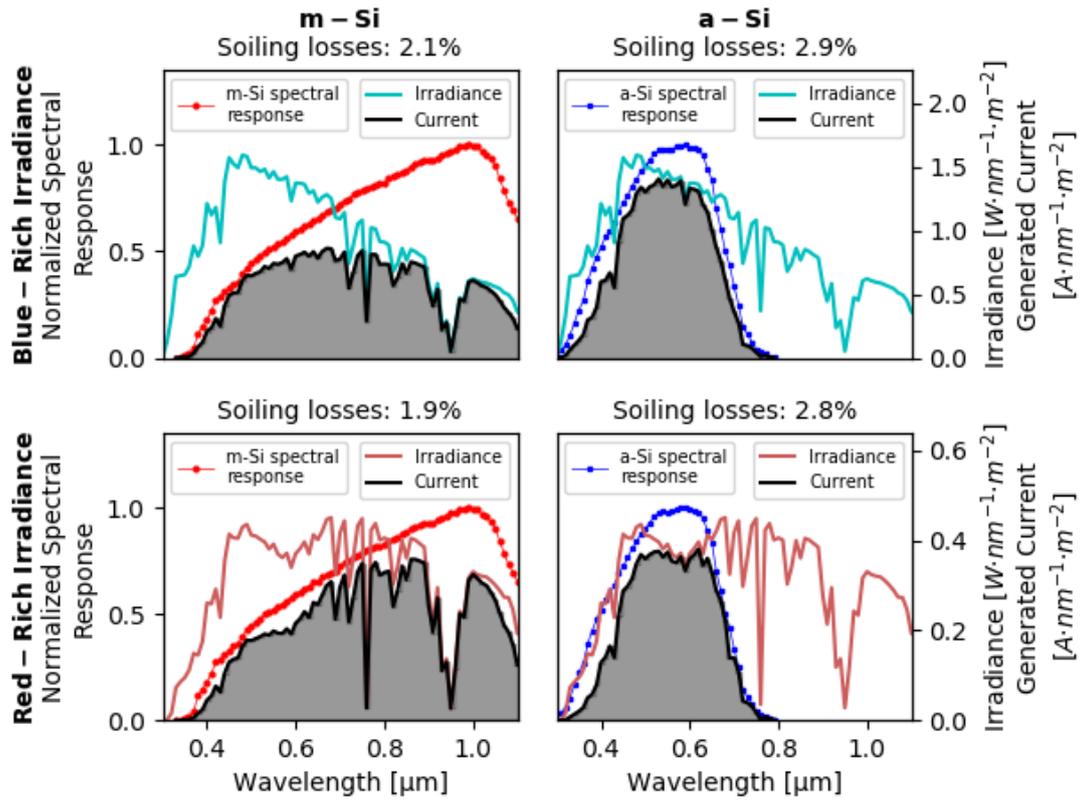

Fig. 2. Soiling losses recorded for Measurement A (shown in Fig. 1, with an average transmittance loss 2.7%) for m-Si and a-Si cells under red-rich or blue-rich irradiance conditions. Each plot shows the spectral response (left axis), the irradiance and the generated current (right axis) for a different combination of irradiance and PV material. The shaded area is the integral in the numerator of Eq. (2).

The aim of this work is investigating methods that can be used to model the full spectral transmittance and that can be used to calculate the soiling loss for PV modules of different materials. These results can find immediate application in the field of soiling monitoring to improve the current optical soiling detection technologies, which are gaining market interest, because of their low cost and good accuracy, but are currently not able to differentiate the impact of soiling on different modules and in different irradiance conditions. The ability to correct the soiling estimation for different PV technologies will also represent an improvement compared to traditional soiling stations [8,17], which are only able to quantify soiling occurring for the PV absorber materials that are used for their soiled and reference modules.

In the next sections, the experimental methods are described that were used to collect and analyse the necessary transmittance data so that a determination can be made using Eq. (2) regarding optimal single and multiple wavelength values for several key photovoltaic technologies.

## 3. Materials and Methods

### 3.1. Transmittance measurements

The same experimental soiling transmittance profiles described previously [3] have been employed in this work. The data was collected from PV glass samples soiled outdoors under national conditions. One Diamant® low-iron glass coupon 4 cm × 4 cm in size and 3 mm thick from Saint-Gobain Glass was mounted from January 2017 to January 2018 on the roof of the A3-building at the University of Jaén (Spain). The coupon was never intentionally cleaned, but rain and dew partially cleaned it from time to time. Its hemispherical transmittance was measured weekly within a wavelength range between 0.300 and 1.240 µm, at 0.0025 µm steps, using a Lambda 950 spectrophotometer with a 60-mm-diameter integrating sphere at the Center of Scientific-Technical Instrumentation (CICT) of the University of Jaén. The technicians at the CICT, who follow the calibration routines indicated by the manufacturer, constantly maintain the spectrophotometer. The weekly hemispherical transmittance of the soiled glass coupon was compared with that of a clean coupon, stored in a dust-free box. All the transmittance profiles due to soiling discussed and shown in this work are obtained as follows, to remove the effect of the glass transmittance:

$$\tau(\lambda) = \frac{\tau_{soil}(\lambda)}{\tau_{ref}(\lambda)} \qquad (3)$$

where $\tau_{soil}(\lambda)$ and $\tau_{ref}(\lambda)$ are the weekly measured spectral transmittances of the outdoor-mounted coupon and of the clean coupon, respectively. The transmittance loss at a given wavelength can be calculated as $1 - \tau(\lambda)$.

The present dataset consists of 34 $\tau(\lambda)$ measurements, collected over 48 weeks. The data shown in this work starts from those measured in the third week of the data collection, here labelled as "week 1". Some weekly measurements are missing because they were not taken, or because the transmittance data was too noisy. Maximum and minimum weekly average soiling transmittances of 1.000 and 0.926 were experienced, with rainfalls and stochastic events affecting the soiling loss profile, as detailed previously [3]. The University of Jaén is located in Jaén (latitude 37º49'N, longitude 3º48'W, elev. 457 m), a high solar insolation location in Southern Spain (> 1800 kWh/m$^2$/year). Extended dry summer seasons, occasional dust storms from the Sahara Desert and periodic burnings of olive tree branches, from extensive local groves, can expose the PV modules to high soiling losses. A description of the weather and soiling conditions experienced during the data collection has been already reported [3,6].

It is important to mention that the value of the average transmittance ($\tau_b$) is strongly affected by the wavelength limits that are adopted in the calculation. Table 1 shows the effect of the selection of the lower limit in the calculation of the average transmittance for Measurement A (shown in Fig. 1 and used in Fig. 2). As can be seen, depending on the lower limit for the value of $\lambda_1$, the average transmittance value might approach the expected soiling ratio for a specific PV technology and move further from the expected soiling ratio for another PV technology. In this work, the average transmittance is calculated for the range 0.3 to 1.1 µm. This is the typical range of spectrophotometers and spectroradiometers.

Table 1. Average transmittance losses, $1 - \tau_b$, for Measurement A (Fig. 1), depending on the minimum wavelength considered. The maximum wavelength is fixed to 1.1 µm. The soiling losses estimated for the given transmittance spectrum are shown in Fig. 2: 1.9-2.1% for a m-Si cell and 2.8-2.9% for an a-Si cell.

| Minimum Wavelength [µm] | 0.300 | 0.325 | 0.350 | 0.375 | 0.400 | 0.425 | 0.450 | 0.475 | 0.500 |
|---|---|---|---|---|---|---|---|---|---|
| Average Transmittance Loss/% | 2.7% | 2.6% | 2.4% | 2.3% | 2.1% | 2.1% | 2.0% | 1.9% | 1.8% |

### 3.2. Spectral transmittance models

The spectral models, which are described below, are compared with three single wavelength value models:

- Broadband (average) Transmittance ($\tau_b$): the transmittance profile is assumed to be flat, with a value equal to the simple average of the spectral transmittance.
- Transmittance at 0.55 μm ($\tau_{0.55}$): the transmittance profile is assumed to be flat, with a value equal to the transmittance measured at a wavelength of 0.55 μm.
- Transmittance at the optimal wavelength ($\tau_{opt}$): the transmittance profile is assumed to be flat, with a value equal to the transmittance measured at the wavelength that returns the lowest error. The optimal wavelength of each material is selected according to the results shown in the previous study [3].

Two multi-variable models have been investigated in this work to replicate the spectral profile of the transmittance due to soiling, and are described in the following sub-sections. The first model is sourced from the literature and, the second one is derived from the latter. Both models make use of one exponential function and of a number of variables. In order to distinguish them in the paper, they have been named according to the number of variables they employ.

### 3.2.1. Three-variable single exponential (3v1e)

The 3v1e equation has recently been utilized to model the spectral transmittance of soiling on glass [2]. The empirical equation, inspired by the Ångström turbidity equation [18], was successfully utilized in the spectral analysis of soiling collected outdoors on PV glass coupons at seven locations worldwide. It is expressed as,

$$\tau(\lambda) = e^{\left(-\beta_{sur}^* \cdot \lambda^{-\alpha^*}\right)} + \gamma^* \tag{4}$$

where $\tau(\lambda)$ is the spectral transmittance at a wavelength $\lambda$ (expressed in μm) and $\alpha^*$, $\beta_{sur}^*$ and $\gamma^*$ are wavelength independent variables. It was suggested that $\beta_{sur}^*$ represents both the mass of particles per unit area on the glass surface and the strength of forward scattering of those particles, $\alpha^*$ relates to the size of the particles and $\gamma^*$ is an offset correction parameter, needed to consider mechanisms taking place when particles are deposited instead of suspended. For example, measurement A in Fig. 1 ($\tau_b$ = 0.973) is fit to $\alpha^*$= 2.954, $\beta_{sur}^*$ = 0.002 and $\gamma^*$= − 0.012, with an R$^2$ = 0.988. Fits for the spectral profiles shown in Fig. 1 are reported in Fig. 3. Counter-intuitively, it was previously found that the analysis of the mass, size or composition of the deposited particles was not required to make use of this empirical approach for natural soiling [2]. The likely reason for this is that while the reflectance of thick layers of dust can exhibit a spectral shape representative of its chemical composition, the forward scattering and transmittance for disperse and isolated particles on a transparent substrate is more strongly dependent on the spectral characteristics of the Mie scattering of the particles. In the analysis that follows, the three variables for the 3v1e method can therefore be determined by fitting the measured transmittance at three distinct wavelengths.

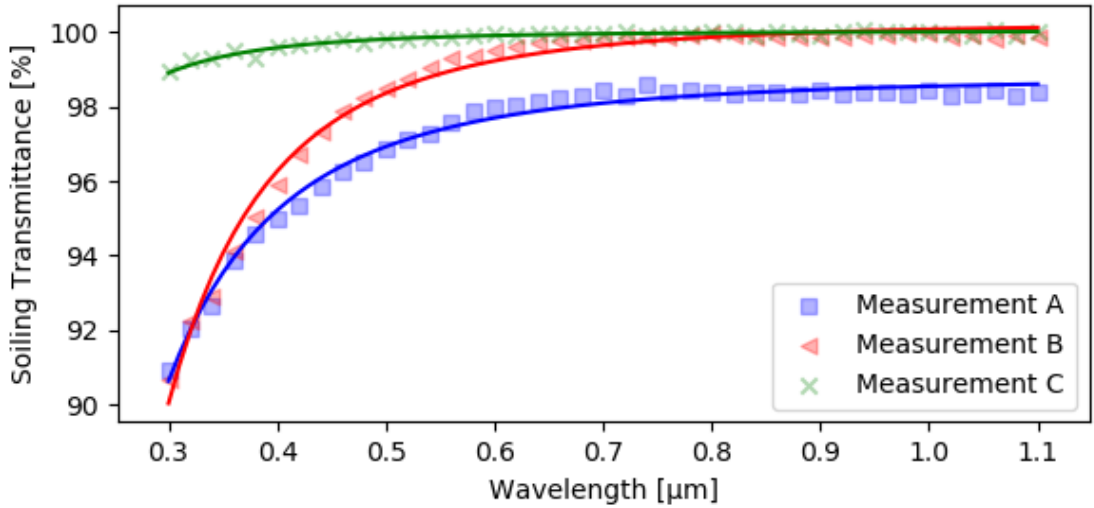

**Fig. 3.** Fits obtained using Equation 4 for the representative hemispherical transmittance spectra shown in Fig. 1.

### 3.2.2. Two-variable single exponential (2v1e)

In the same paper [2], direct correlations between the variables were found. One connected $\beta^*_{sur}$ to the broadband (average) transmittance, $\tau_b$ ($R^2$ = 0.99). Another correlation was between $\gamma^*$ and $\tau_b$ ($R^2$ > 0.99). These are,

$$\tau_b = -10.99 \cdot \beta^*_{sur} + 1.01 \tag{5}$$
$$\tau_b = 1.30 \cdot \gamma^* + 1.00 \tag{6}$$

This finding seems to suggest that $\beta^*_{sur}$ and $\gamma^*$ could be correlated, reducing the number of input parameters from three to two. A combined equation can therefore be written as,

$$\tau(\lambda) = e^{\left(-\beta^*_{sur} \cdot \lambda^{-\alpha^*}\right)} - 8.45 \cdot \beta^*_{sur} + 0.01 \tag{7}$$

where the term $\gamma^*$ is replaced with the expression $-8.45 \cdot \beta^*_{sur} + 0.01$. Compared to the previous model 3v1e, this equation requires one less variable. The same correlation, with a slight difference in the value of the slope, can be found if the $\beta^*_{sur}$ and $\gamma^*$ data from Table 2 of previous work [2] are directly plotted, as shown in Fig. 4. Analogous to the 3v1e case, $\tau(\lambda)$ is the spectral transmittance at a wavelength $\lambda$ (expressed in μm) and $\alpha^*$ and $\beta^*_{sur}$ are wavelength independent variables. It should be noted that the range for the values in Fig. 4 is quite large and represents both high and low soiling locations. For the 2v1e work, the relationship in Eqn. (7), and results of Fig. 4, allows for the determination of the values for all of the variables by fitting the transmittance data at two distinct wavelengths.

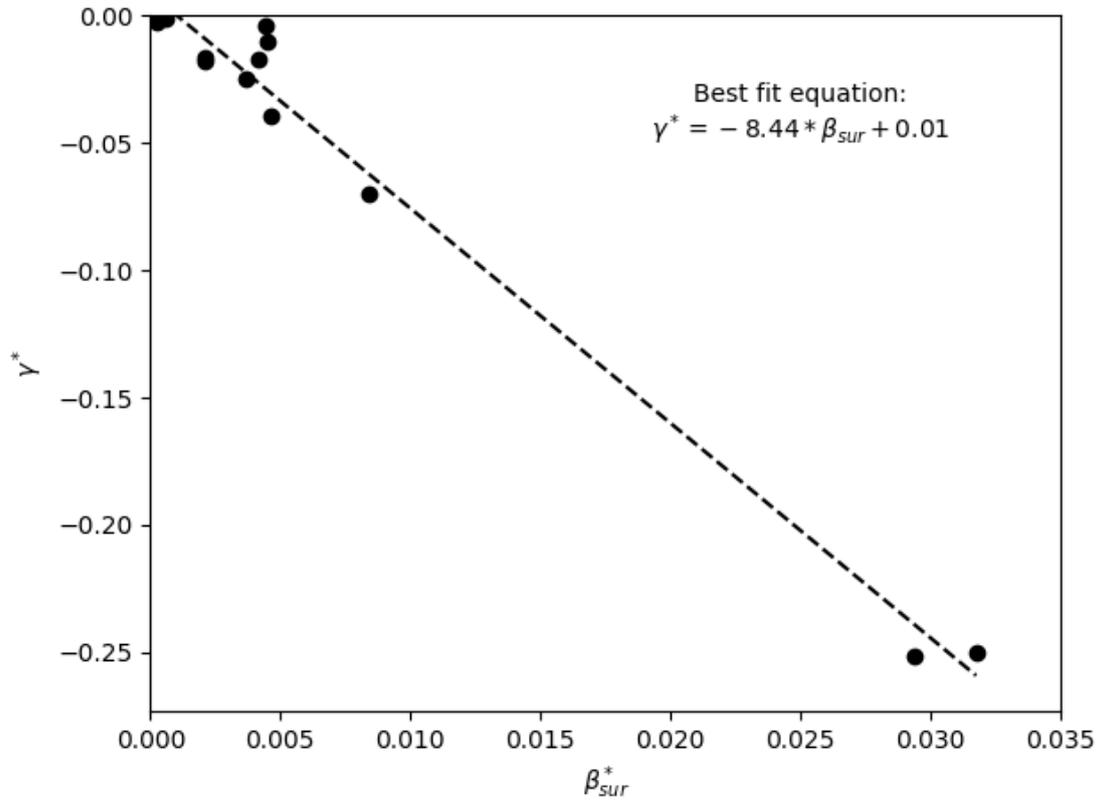

Fig. 4. The correlation between $\beta^*_{sur}$ and $\gamma^*$ from Table 2 of prior work that examined soiling at 7 locations worldwide [2].

### 3.3. Irradiance, Spectral Response and Soiling Ratio

As in previous work [3], six representative PV materials have been considered. Their spectral response profiles are shown in the upper plot of Fig. 5. These are examples of spectral response curves for the different PV materials, and were sourced from previous studies [19–21]. In the bottom plot of the same figure, the three spectral irradiances considered in our analysis are shown. These have been selected to represent different conditions: the AM1.5 global irradiance, reference, spectrum (ASTM G173-03 standard), a blue-rich spectrum and a red-rich spectrum. The reference spectrum has been sourced from an open source database [22], while the last two spectra have been generated through the SMARTS radiative transfer model [23]. The IEC 61724-1 standard that covers soiling measurements states the test should be performed at during ± 2 hour window around local solar noon for fixed systems and at times when the angle of incidence is < 35 ° for tracked systems [13]. Given the atmospheric conditions found at a wide range of latitudes and seasons, one can expect that the solar spectra incident on a PV module will vary widely for soiling measurements in the field.

The variables provided as input to the model to generate the irradiance profiles are shown in Table 2, along with each spectrum's Average Photon Energy (APE) [24,25]. This index describes the sunlight's chromatic distribution, with higher APE values corresponding to "blue-richer" spectra while lower APE values corresponding to "red-rich" spectra [26]. In this study, the APE has been calculated for wavelengths between 0.3 and 1.1 μm, the same waveband employed for all the spectra investigated in this study, at steps of 0.01 μm. The whole analysis is conducted considering the PV modules at a fixed reference temperature.

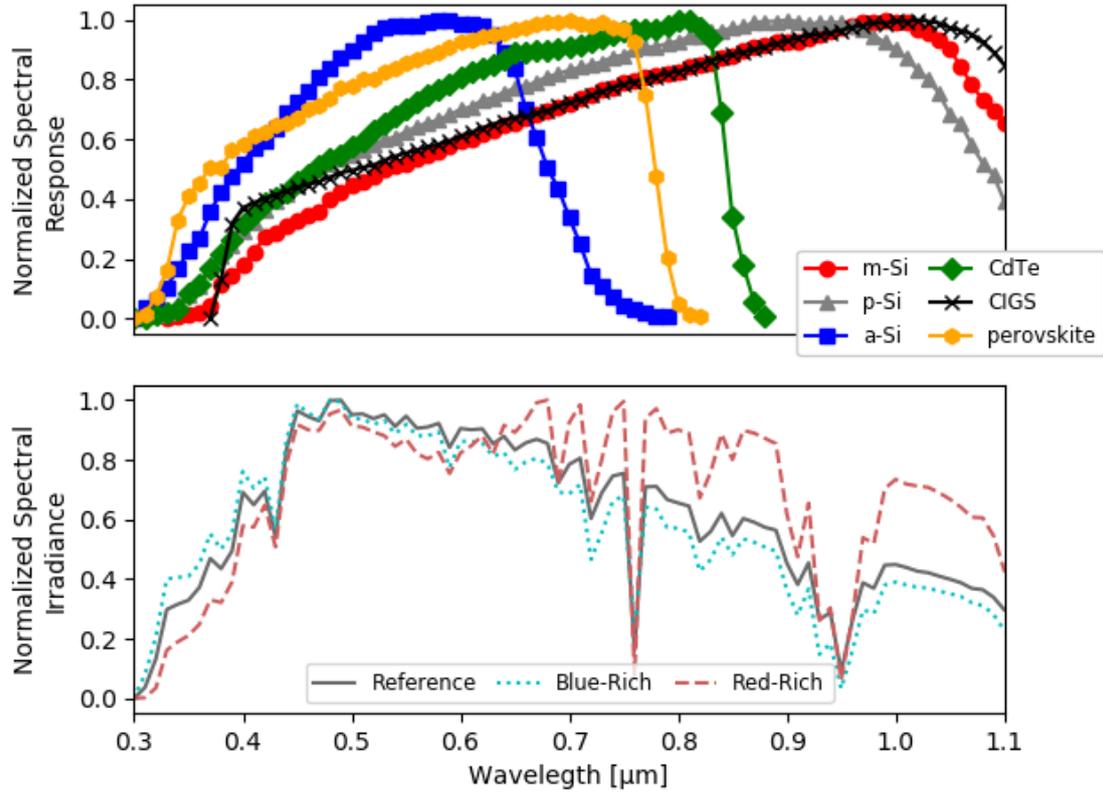

**Fig. 5.** Normalized Spectral Response of the six PV materials (top chart) and the normalized profiles of the three irradiances (bottom chart) considered in this work. The normalized spectral responses are representative for each material and were sourced from previous studies [19–21].

**Table 2.** Characteristics of the irradiance spectra used in this work. The Average Photon Energy has been calculated between 0.3 µm and 1.1 µm at 0.01 µm steps.

| Irradiance | Air Mass | Aerosol Optical Depth at 0.5 µm [cm] | Precipitable Water [cm] | Average Photon Energy [eV] |
|---|---|---|---|---|
| Reference Spectrum | 1.5 | 0.084 | 1.42 | 1.85 |
| Blue-Rich | 1.0 | 0.100 | 4.00 | 1.91 |
| Red-Rich | 5.0 | 0.400 | 1.25 | 1.74 |

### 3.4. Metrics for the Evaluation of the Fit to Models

The curve fitting has been performed through the *curve_fit* function in the *SciPy* library for Python 2.7 [27], which uses nonlinear least-squares with a Trust Region Reflective algorithm. The initial guesses and the boundary conditions for each variables were set according to those reported previously [2]. The maximum number of iterations allowed for fitting was 100,000. The quality of the models investigated in this work has been assessed using the following indexes:

- The mean absolute error (MAE) expresses the average value of the absolute errors between the measured and modelled data. It is 0.0% if the modelled data have the same values of the measured data; otherwise, it rises depending on the number and the magnitude of the errors in the prediction. The MAE, expressed as a percentage, is obtained as,

$$\text{MAE } /\% = \frac{100}{n}\sum_{i=1}^{n}\left|Z_{mod,i} - Z_{meas,i}\right| \tag{8}$$

where $Z_{mod,i}$ and $Z_{meas,i}$ are the $i^{th}$-pair of modelled and measured data, and *n* is the total number of pairs.

- The mean error (ME) expresses the average value of the errors between the measured and modelled data. It provides information on the systematic bias in the models; it is positive or negative depending if the modelled data, respectively, overestimates or underestimates the values of measured data. A ME of zero is due to the lack of a systematic bias, but does not necessarily express a perfect correlation between measured and modelled data. The ME is obtained as,

$$\text{ME} / \% = \frac{100}{n} \sum_{i=1}^{n} (Z_{mod,i} - Z_{meas,i}) \quad (9)$$

## 4. Spectral Transmittance Modelling

### 4.1. Performance of the Models

The models listed in Section 3.2 have been tested to assess their ability to reproduce the spectral profiles of the 34 transmittance measurements taken during the data collection period. For each week, a number of simulated transmittance profiles were generated by using both the flat and the spectral models and these were then compared with the measured transmittance spectrum due to soiling. The errors found in the estimation of each week's spectral transmittance profiles were calculated from Eqs. (8) and (9) and are shown in Fig. 6.

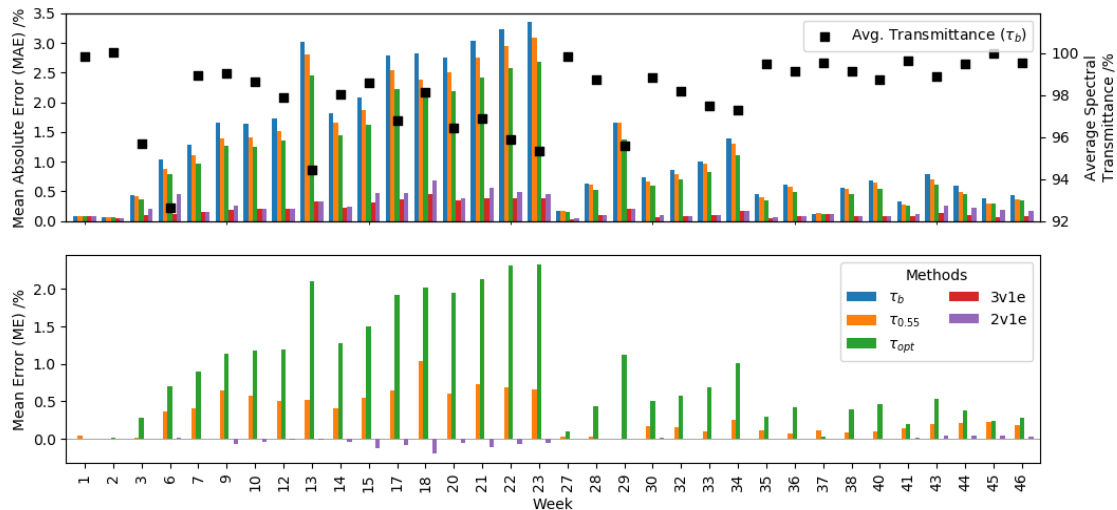

**Fig. 6.** MAE and ME comparing the measured spectral transmittance with the flat transmittance profiles modelled using the average transmittance ($\tau_{avg}$), with the transmittance at 0.55 μm ($\tau_{0.55}$) and the transmittance at the optimal wavelength ($\tau_{opt}$) approaches, and with the spectral transmittances modelled using the three-variable single exponential (3v1e), and the two-variable single exponential (2v1e) methods.

As expected, using a spectral model can significantly reduce the error in the estimation of the spectral transmittance. In particular, the three-variable single exponential spectral model is found to consistently return the best correlations, with a maximum MAE below 0.5%, for hemispherical transmittance losses ranging between 0.0% and 7.4%. Using two variables instead of three produces slightly larger errors, up to 0.7% MAE.

On the other hand, assuming a flat transmittance profile can produce significant errors (up to 3.4% MAE), especially in conditions of higher soiling. For the given dataset, a single wavelength measurement returns lower absolute errors than the average transmittance value, even if it is exposed to a systematic over-estimation of the transmittance (and under-estimation of the losses) as proved by the high positive ME values. The optimal wavelengths range between 0.6 and 0.8 µm, in the region of the spectra where the losses are at a minimum. It is interesting to mention that this approach limits the MAE, but produces the largest ME (i.e. the largest over-estimation of the transmittance and the largest under-estimation of the losses).

These aspects can be explored further by looking at the errors for the transmittance data from each week, and how this depends on the characteristics of the transmittance spectra. This is shown in Figure 5. The minimum transmittance in the first row represents the minimum value that each weekly spectrum reaches (typically near 0.3 µm). From the first row of Fig. 7, we see that the MAE values for the all the methods are found to be the highest for those transmittance spectra that have the lowest minimum transmittance values. This means that, among the spectral transmittance profiles shown in Fig. 1, weeks 10 (B in Fig. 1) and 34 (A in Fig. 1) record the highest MAEs (with minimum single value transmittances of 0.907 and 0.909, respectively), while week 27 (C in Fig. 1) returns the lowest MAEs (with a minimum single value transmittance of 0.988). This correlation between error and minimum transmittance value becomes more significant for the flat models (first three columns of Fig. 7); in these cases, the error varies by one order of magnitude more than for the spectral models. This is due to the fact that flat models do not take into account the increased amount of losses occurring in the blue region.

This result is confirmed if, for each transmittance spectrum in the dataset, the average spectral transmittance of each waveband (UV, visible and near infrared) is compared to the broadband transmittance (i.e. the average transmittance across the whole spectrum). This is done through the Waveband Specific Transmittance (WST), which expresses the ratio between the average spectral transmittance of a waveband $i$ and the broadband transmittance, calculated as,

$$WST_i = \frac{\int_{\lambda_{1i}}^{\lambda_{2i}} \tau(\lambda) \cdot d\lambda / (\lambda_{2i} - \lambda_{1i})}{\int_{0.3\ \mu m}^{1.1\ \mu m} \tau(\lambda) \cdot d\lambda / (1.1\ \mu m - 0.3\ \mu m)} \tag{10}$$

where $\lambda_{1i}$ and $\lambda_{2i}$ are the shortest and longest wavelengths of the $i$-th waveband. According to its definition, it has a value < 1 (< 100%) if the transmittance of the $i$-th waveband is lower than the broadband transmittance and has values > 100% otherwise. The smaller its value, therefore, the more the losses are in the specific waveband. For the analysed dataset, the plots reported in the last three rows of Fig. 7 show that for more losses incurred in the UV (i.e. the smaller $WST_{UV}$), the larger the mean absolute errors, especially for flat transmittance profile approaches. On the other hand, for more losses incurred in the near infrared (i.e. the smaller $WST_{NIR}$), lower errors are expected. One should note that for the last two columns of Fig. 5, the x-axis scale decreases dramatically.

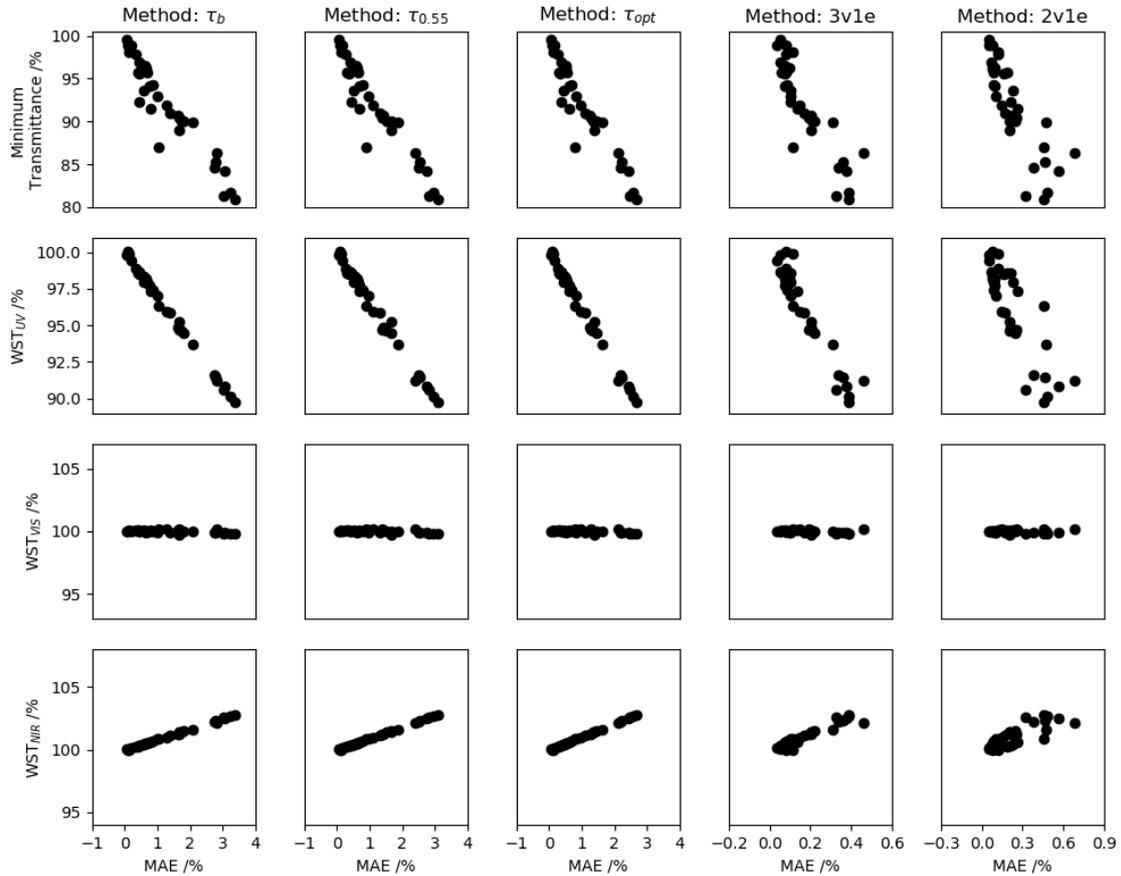

**Fig. 7.** MAE returned by each model depending on the characteristics of the transmittance spectra for each week of data collection. The minimum transmittance represents the minimum value that each weekly spectrum reaches. The Waveband Specific Transmittances in the Ultraviolet ($WST_{UV}$), in the Visible ($WST_{VIS}$) and Near Infrared ($WST_{NIR}$) are calculated by using Eq. (10), according to the waveband limits reported in the prior study [3]. Plots on the same row share the same y-axis, and plots on the same column share the same x-axis.

### 4.2. Optimal combinations of measurements

In the previous section, we have shown how the use of spectral models can improve their utility as replacements for the full, measured soiling transmittance profiles. These models are based on two or three variables, which means that the full transmittance spectrum for soiling can be modelled by measuring the transmittance at only two or three wavelengths, respectively. In this section, we aim to understand which combinations of wavelengths can be used to model the transmittance spectra due to soiling with the highest accuracy. As a first step, each week's transmittance profile has been modelled by using the transmittance measured at a number of wavelengths equal to the number of each model's variables. Wavelengths between 0.30 μm and 1.10 μm have been considered, at steps of 0.05 μm. The results from all of the transmission curves of the study were combined. The results show the mean MAEs, obtained as an average of the MAEs calculated for each of the 34 transmittance curves, following these steps:

1. Each transmittance curve is modelled and a MAE is calculated.
2. The average of the MAEs is calculated.

Fig. 8 shows the average MAEs when two transmittance values are provided to the two-variable single exponential (2v1e) model. The lowest errors are found if the transmittance is measured at 0.35 μm for one of the wavelengths and between 0.7 μm and 1.0 μm for the second wavelength. In general, the worst results are found for combinations of wavelengths where both are greater than or equal to 0.70 μm. Also, the use of consecutive wavelengths (e.g. 0.35 μm and 0.40 μm or 0.50 μm and 0.55 μm) should be avoided.

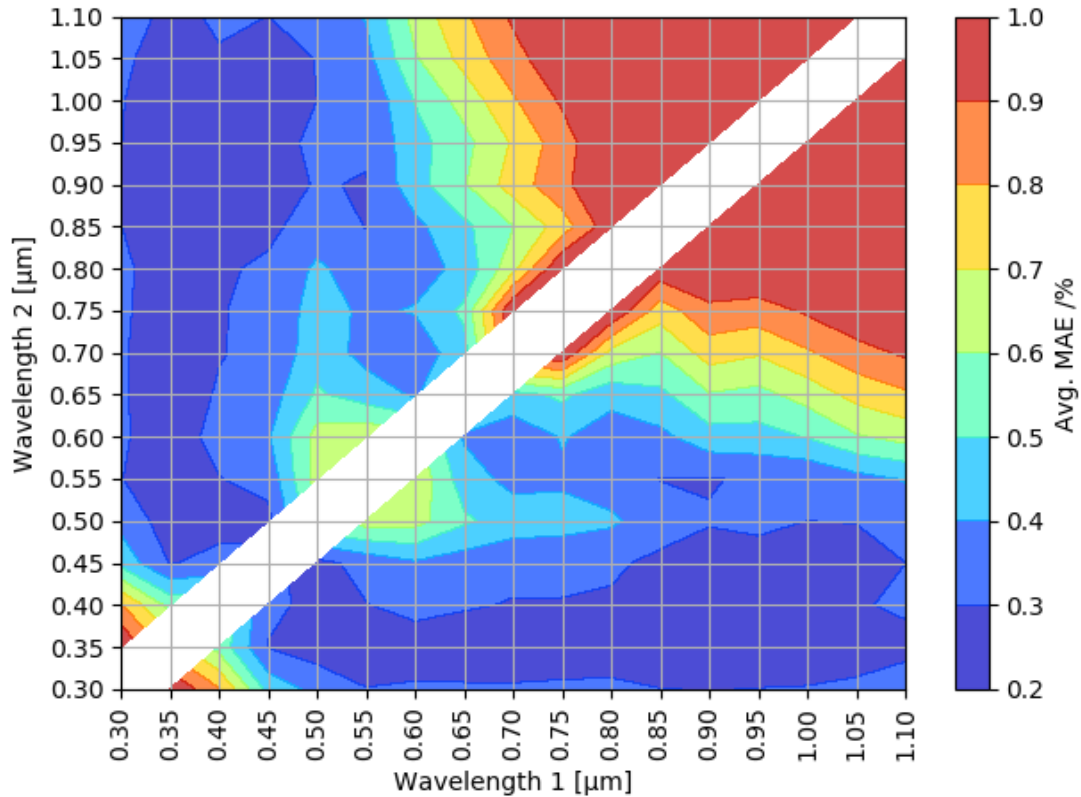

Fig. 8. Average MAE for the two-variable single exponential (2v1e) model over the various weeks of the experimental investigation. These considered two wavelengths as input. The contour plot is created from a 0.05 µm x 0.05 µm grid and is mirrored on the diagonal. Combinations where the two wavelengths are the same are shown in white. The best result is for a wavelength pair of 0.35 µm and 0.85 µm. Any MAE ≥ 1.0% is coloured in red.

For the three-variable single exponential (3v1e) model, it was found that all but four weeks yielded a fit to Eqn. 4 with an $R^2$ > 0.95, thus further validating our empirical approach. In the analysis that follows for 3v1e, three wavelengths were provided for the fit to that equation. The average MAEs found for each possible combination are shown in Fig. 9. In this case, the minimum MAE can be lowered to less than 0.2%, with the best results returned for a combination of these wavelengths: 0.35 µm, 0.45-0.50 µm and 0.85-0.95 µm. A measurement at 0.35 µm is still found to be essential to minimize the error, because it makes it possible to correctly model the attenuation in the UV. The addition of an intermediate wavelength makes it possible to improve the fit, because of the noticeable drop in transmittance from the green to the blue and UV regions (as shown in Fig. 1). Two value combinations of low and medium wavelengths (≤ 0.70 µm), as well as combinations of medium and high wavelengths (≥ 0.50 µm), should be avoided.

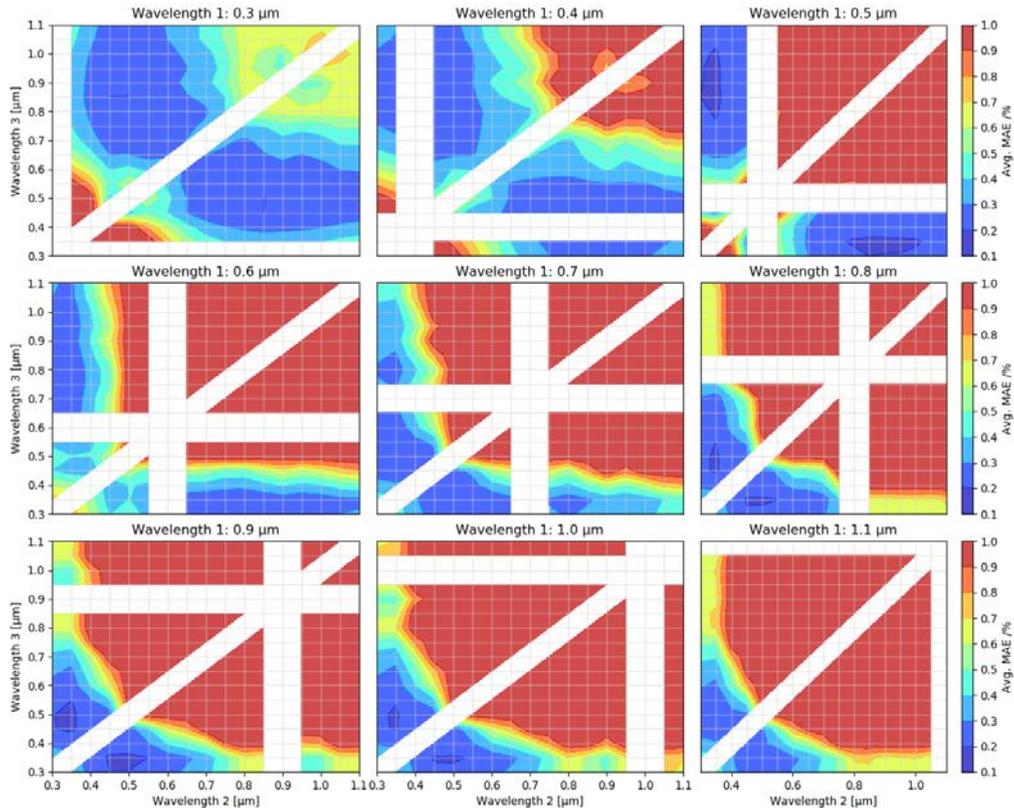

**Fig. 9.** Average MAE for the three-variable single exponential (3v1e) model over the various weeks of the experimental investigation. These considered three wavelengths as input. The contour plots are created from 0.05 µm x 0.05 µm grids and are mirrored on the diagonal. Combinations where at least two wavelengths are the same are shown in white. As one can see from the deeper blue region of the 0.5 µm (upper right) plot, the best result is for the combination: 0.35 µm, 0.50 µm, and 0.85 µm. Any MAE ≥ 1.0 % is coloured in red.

## 5. Optical Detection of the Soiling Ratio

### 5.1. Soiling Ratio Estimation

In our previous work [3], it was shown that is possible to estimate the soiling losses from single value transmittance measurements and it was also concluded that each PV material had a specific wavelength at which the error was minimized. In this work, we aim to further lower the error in the soiling estimation, by modelling the full soiling transmittance spectrum through a limited number of measurements. In the previous section we proved how it is possible to model a full transmittance spectrum using a limited number of single-value measurements (≤ 3). Using Eq. 2, we can then use these spectra to estimate the soiling losses incurred by different PV materials under several irradiance conditions.

Fig. 10 shows the error in the estimation of the soiling ratio when each model is used for different PV materials under three different irradiance conditions. As can be seen, the largest errors are found if a constant transmittance is assumed. The magnitude of the error is dependent on the spectral distribution of the irradiance. The errors are higher for lower energy gap materials (Si, CdTe) under conditions of red-rich spectra, where their spectral response is at maximum, and, similarly, the errors are larger for larger energy gap materials (a-Si, perovskite) for conditions of blue-rich spectra. This means that under conditions of favourable spectral irradiance for each material, the error in the estimation of the soiling ratio is at maximum.

On the other hand, the spectral models always return among the lowest errors, independently of the conditions. Indeed, the spectral transmittance models are found to be the more robust to the change in irradiance spectrum. The use of the spectral transmittance models is found to be particularly important for low and intermediate energy materials, because of their extended absorption band, which goes from the strongly affected blue wavelengths to the less impacted longer wavelengths. It should be noted that the plots for p-Si and CIGS are not shown, as the results are similar to those found for m-Si, since they all have very similar small bandgaps.

As was discussed in connection with the first two rows of Fig. 7, the errors are found to increase with the severity of soiling. Comparing the left and right columns of Fig. 10, noting the change in the x-axis scales, this can also be seen. This suggests that the errors in soiling ratio predictions based on a flat transmittance assumption (using a single-value measurement) can also be erroneous in locations or seasons with extreme soiling.

Our previous work [2] analysed the data obtained from glass coupons after eight weeks of soiling deposited outdoors at seven sites worldwide. We found that the shape of the relative transmittance spectra fit Eq. (4) and the 3v1e approach quite well. That finding gives us some confidence in hypothesizing that the conclusions from Fig. 10 can be extended to other sites. Also reported in the prior work was that the broadband transmission ($\tau_b$) for a soiled silicon PV device is equal to its predicted soiling ratio, with an error < 0.7%. While that is consistent with Fig. 10, the present study extends the result with a more extensive dataset from almost a year of outdoor exposure at a single site. It also strongly suggests caution in the use of single wavelength values, compared to the accuracy in the estimates for $r_s$ that can be achieved using spectral models such as 2v1e and 3v1e.

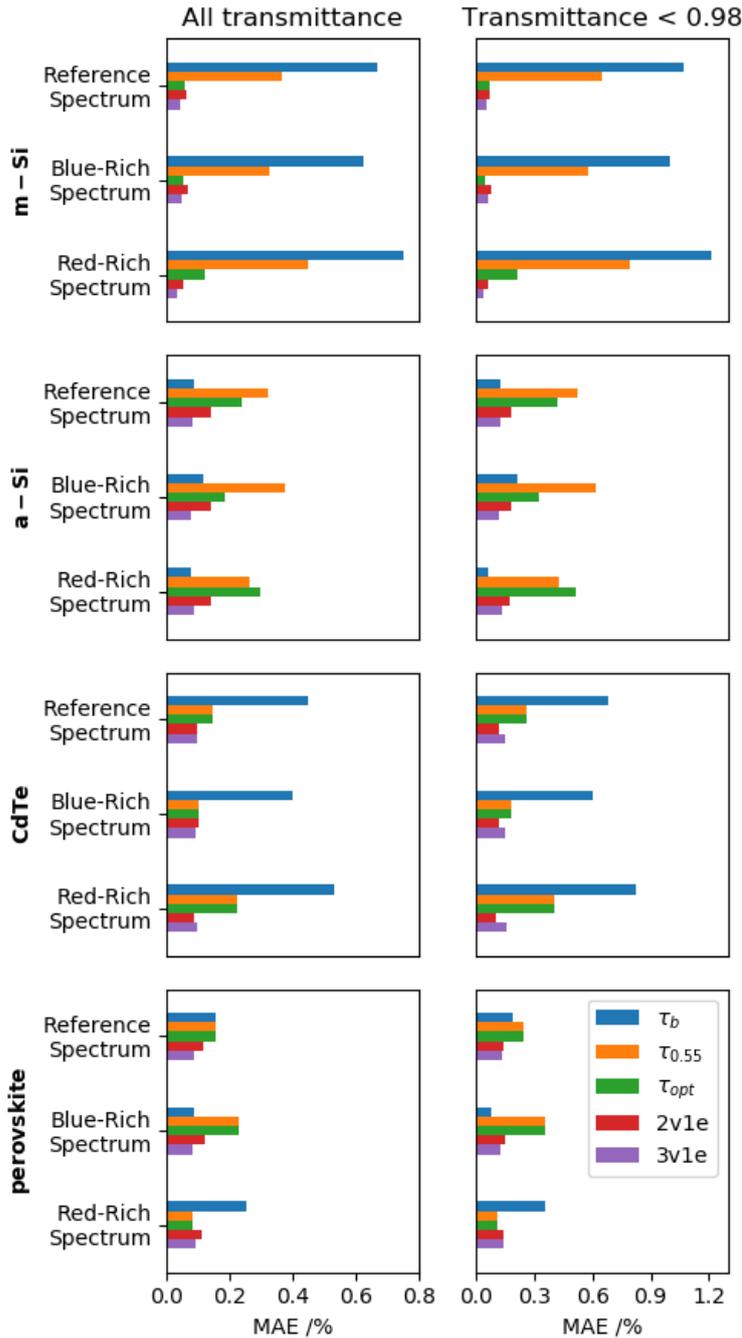

**Fig. 10.** MAE in the calculation of the soiling ratio, $r_s$ for the 46-week dataset, depending on the PV material, the irradiance, and the spectral modelling method used for $\tau$. The optimal wavelengths recommended in the previous study [3] have been used for the $\tau_{opt}$ method. The plots on each column have the same x-axis scale. Similar results are obtained for m-Si, p-Si and CIGS. For better readability, only the m-Si plots are shown.

### 5.2. Specific optimization for key PV technologies

In the previous section, we showed which combinations returned the lowest errors when the three-variable single exponential (3v1e) model was employed. In this section, we want to investigate if it is possible to further lower that error by using the two-variable equation and by tuning the wavelengths according to the PV material in use. Indeed, each PV material is able to work in a different absorption band. This means that the estimation of soiling might rely on modelling only a limited portion of the spectrum instead of the full spectrum. Therefore, it might be possible to identify combinations of wavelengths that optimize the soiling ratio estimation for the spectral absorption of a selected PV material. For this reason, the analysis described in the previous section has been repeated by taking into account each material's specific absorption band and the three irradiance spectra considered in this work. The results, plotted in Fig. 11, confirms that each material has a set wavelength that can optimize the modelling and that the combination varies according to the PV material and its spectral response. The combinations are found to only slightly vary depending on the characteristics of the spectral distribution of irradiance.

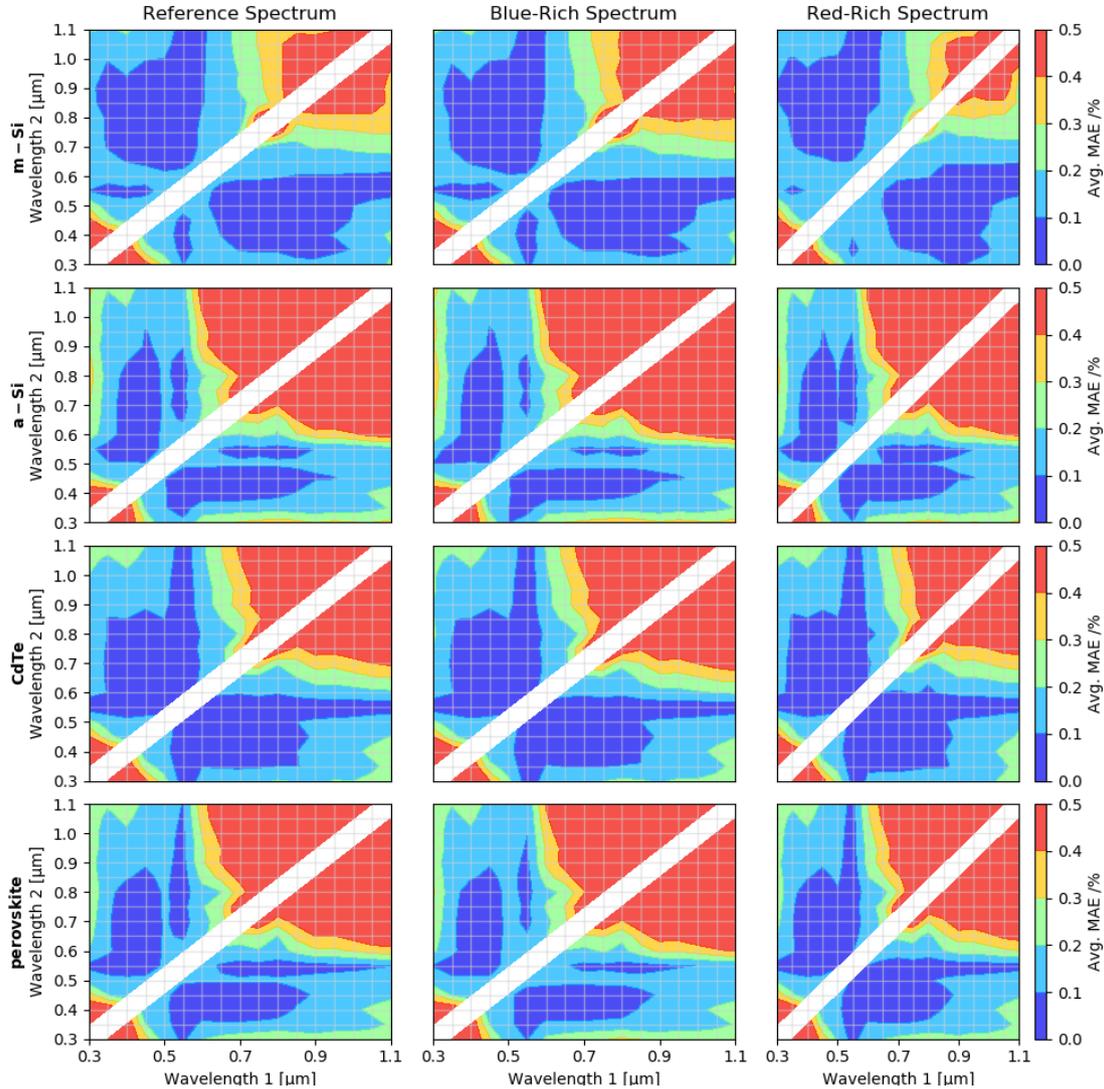

**Fig. 11.** Average MAE to estimate the soiling ratio for various PV materials and irradiance conditions, when the two-variable single exponential (2v1e) model is employed and the transmittance values at two indicated wavelengths are provided. Plots for p-Si and CIGS are similar to m-Si and are not shown.

Fig. 11 displays a summed error for all of the two wavelength combinations used in the 2v1e spectral model. For each particular PV material, it is possible to identify from our 34-week dataset at least one combination of two wavelengths that reduces the error that is found to a level lower than the case for the three-variable equation (3v1e). This was done for all three of the investigated irradiance conditions using the insights of Fig. 10. Within the results shown in Fig. 12, there are some noteworthy findings. PV absorber materials CdTe, amorphous silicon and perovskite have a common area of roughly rectangular shape. The optimal combinations obviously depend on the absorption characteristics of each PV material; amorphous silicon and perovskite require only low or intermediate wavelength combinations (≤ 0.8-0.9 μm), while low bandgap energy materials like silicon require combinations of 0.55 µm and an IR wavelength (≥ 0.85 µm). It should be pointed out that the classical Ångström turbidity equation is routinely employed in atmospheric studies by using measurements from a Sun Photometer at two or three carefully selected wavelengths [28]. That equation formed the basis for Eq. 4 [2].

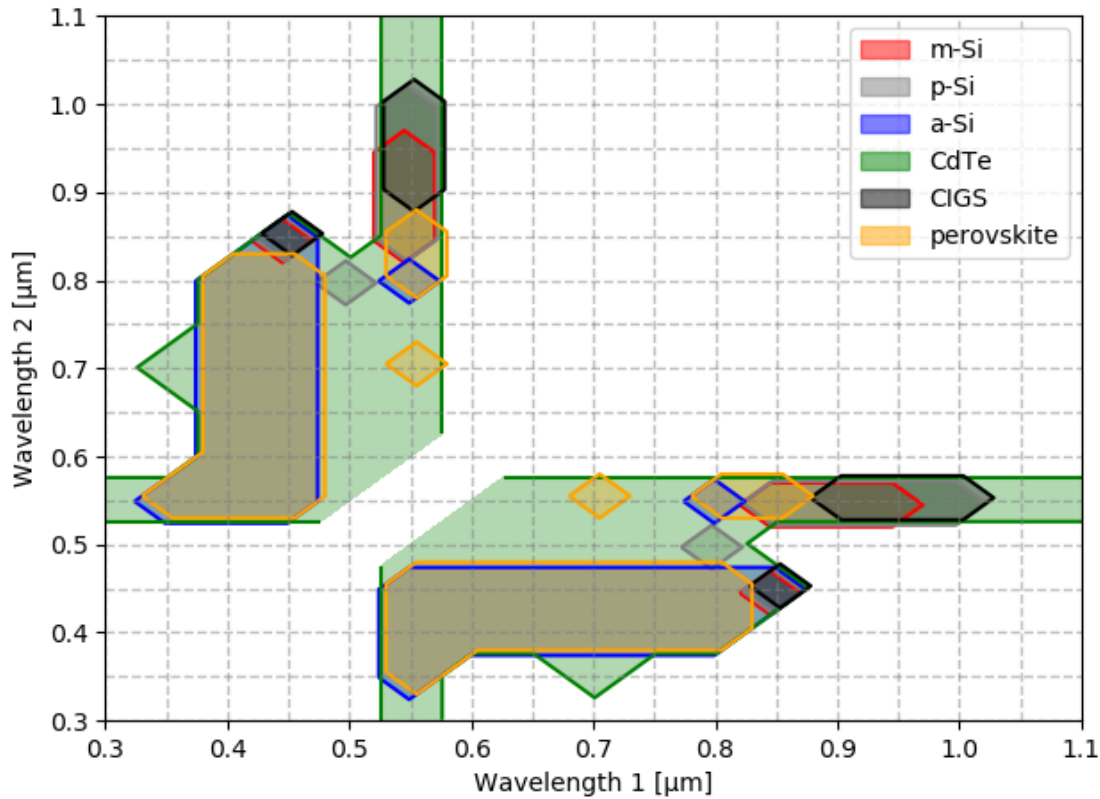

**Fig. 12.** PV material's optimized wavelength combinations that produce errors in the estimation of the soiling ratio for the two-variable single exponential (2v1e) model that are lower than the errors found for the three-variable single exponential (3v1e) under the three irradiance conditions considered in this work. Each coloured area groups the combinations of a specific material. The figure is created from a 0.05 µm x 0.05 µm grid.

### 5.3. Robustness of the models versus measurement uncertainty

So far, the present analysis has not taken into account how a transmittance measurement error can propagate in the estimation of the soiling ratio. In general, a ±0.5% uncertainty at each wavelength is associated with the transmittance measurements made with a spectrophotometer [2]. The aim of this section is to quantify the robustness of each of the investigated methodologies to a random measurement error. For this reason, the previous analysis has been repeated for 1000 iterations, by adding a random error between -0.5% and +0.5% to the transmittance data for each wavelength. For each iteration, the difference between the modelled average transmittance, where the models have had input data subject to the artificial perturbation errors, and the original average weekly transmittance has been calculated.

Fig. 13 shows the results of this analysis, in terms of average error over the 46 weeks of investigation, iterated 1000 times each, under different irradiance conditions and for different PV materials. The errors are calculated as the absolute difference between the soiling ratio modelled from the artificial (random, erroneous) spectra, and the soiling ratio as calculated by using all the data points of the original non-manipulated transmission spectra. The spectral models return the lowest errors in most of the cases, compared to the flat transmittance profile approaches. This is particularly clear for low and intermediate energy gap materials, for which a single measurement is not enough to model the spectral behaviours over their extended absorption wavebands.

On the other hand, the differences are similar for most of the models when a-Si and perovskite materials are considered; this is due to the limited range of spectral response (waveband) of these higher bandgap materials. In these conditions, the average transmittance is found to provide one of the best results, because the error is attenuated thanks to the larger number of data points considered with this approach.

In general, the performance of the spectral models tends to improve under conditions of the reference and red-rich spectra. Overall, the 2v1e model, when fed with measurements at PV material-optimized wavelengths, is found to return the lowest errors in all the cases. It is important to mention that this 2v1e* approach would optimize the performance for a single PV material only, as different wavelength combinations are required for each of the various PV materials. Otherwise, if multiple PV technologies need to be investigated with one soiling sensor, the other two spectral models should be utilized.

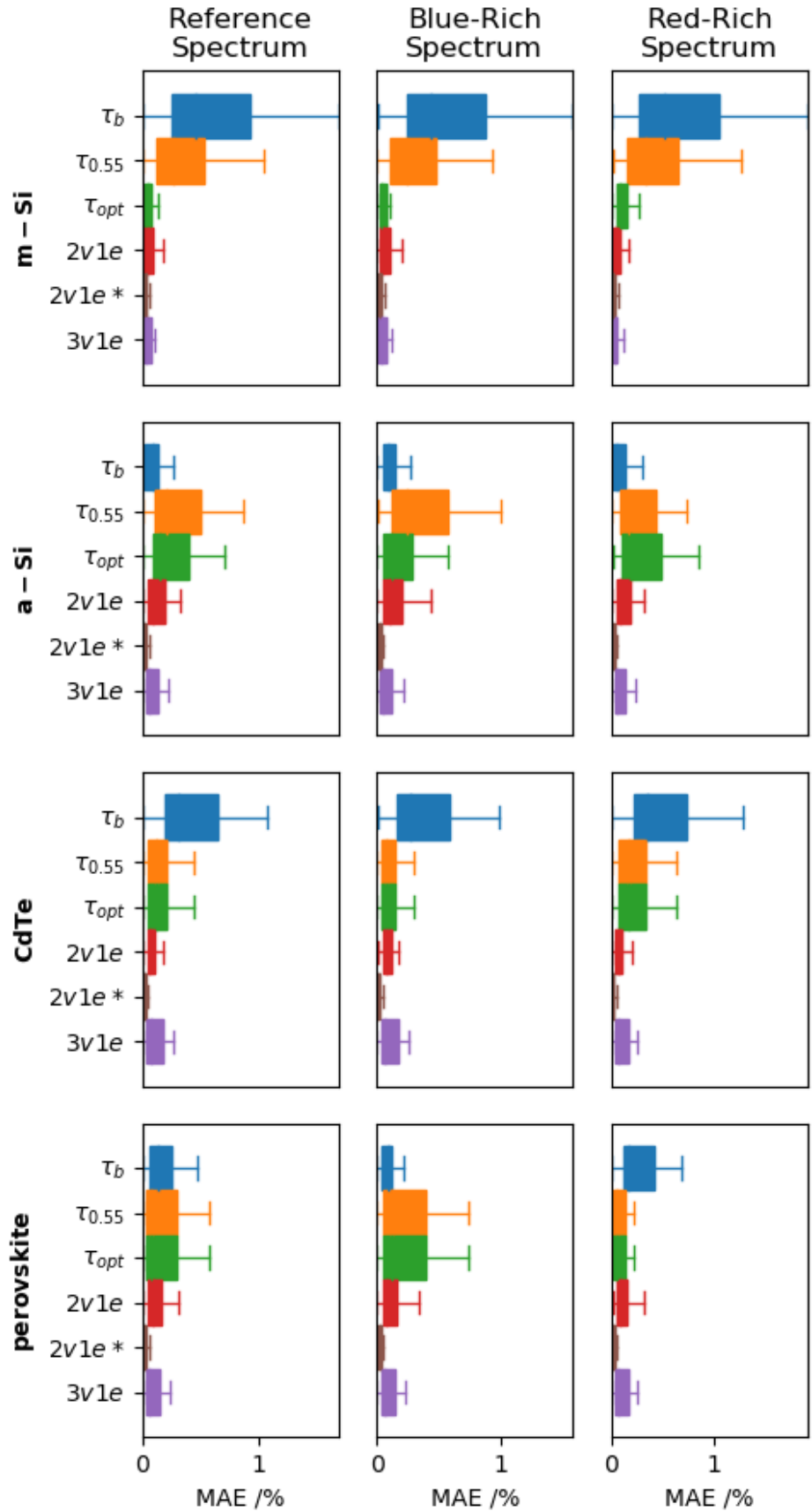

**Fig. 13.** Boxplots representing the mean absolute errors (MAE) found after 1000 iterations in which the soiling ratios, $r_s$, are modelled by introducing a random error of ±0.5%, at each wavelength, for the transmittance spectra used as input, compared to the soiling ratios calculated from the original spectra. The 2v1e and the 3v1e

approaches use the same wavelengths for all the PV materials. The 2v1e* method uses optimized combinations of wavelengths for each PV material. Several key PV materials and three irradiance conditions are indicated. Outlier points are not shown. The plots on each row have the same y-axis scale. Plots for p-Si and CIGS are not shown.

### 5.4. Errors for Larger Soiling Ratios

We turn our attention back to Figs. 1 and 2, to examine the errors in soiling ratio estimates when single wavelength approaches are used. As discussed in the presentation of Fig. 5 and Fig. 8, these are expected to grow as the broadband (average) transmittance loss, $(1 - \tau_b)$, due to soiling increases. As a specific example, the analysis shown in Fig. 2 can be applied to week 34, case A shown in Fig. 1, and week 23, which has a higher average transmittance loss. The results are summarized in Table 3. One can see that using the optimal wavelength [3] for m-Si (0.6 μm) is reasonable for both weeks. One can also see for the same PV material (m-Si) that the error between $(1 - \tau_{opt})$ and the soiling loss values in the table can indeed be larger when the transmission losses are greater. Also apparent is that the use of the same wavelength for another material (a-Si) is not warranted and that, again, the absolute difference in soiling loss between blue and red irradiance is larger for the case of heavier soiling.

It is therefore clear from this example – and from the analysis presented throughout this work – that the use of spectral models and the techniques which we have applied to a particular location in southern Spain should be repeated for other sites with diverse climates and varied soiling conditions. Concurrently, the empirical relationships of Eqs. (4)-(6) should also be further validated using other datasets. In addition, further experimental work should be carried out to correlate the estimated soiling losses to those measured using PV modules in the field. These studies can lead to the creation of better monitoring equipment to establish power losses due to soiling.

Table 3. Soiling losses, $1 - r_s(t)$, measured in weeks 34 and 23 for m-Si and a-Si under red-rich and blue-rich irradiance conditions.

| WEEK 34<br>Average Transmittance Loss: 2.7%<br>Transmittance Loss at 0.6 μm: 2.0% | | | WEEK 23<br>Average Transmittance Loss: 4.7%<br>Transmittance Loss at 0.6 μm: 3.1% | | |
|---|---|---|---|---|---|
|  | m-Si | a-Si |  | m-Si | a-Si |
| Blue-Rich Irradiance | 2.1% | 2.9% | Blue-Rich Irradiance | 3.1% | 5.1% |
| Red-Rich Irradiance | 1.9% | 2.8% | Red-Rich Irradiance | 2.8% | 4.8% |

### 6. Conclusions

Monitoring soiling is an essential task to minimize the losses of PV systems deployed worldwide, while limiting the operation and maintenance costs. In order to limit the installation and operation costs, the soiling sensor market has been moving toward maintenance-free optical soiling detectors that measure the transmittance of soiling to quantify the soiling losses occurring on a PV system. Some commercial soiling sensors currently available are based on a single optical measurement, which can give a good estimate of the soiling trends, but do not consider the spectral profile of soiling and do not perform any correction based on the spectral irradiance distribution or the spectral response of the PV materials they are monitoring. In particular, the errors of single-measurement systems are expected to rise with the severity of soiling, due to the larger soiling loss occurring at shorter wavelengths.

In this work, we investigated the possibility of estimating the soiling transmittance spectra using empirical spectral models and compared these to the use of single wavelength transmittance measurements. For the two spectral models studied, we identified optimal wavelength combinations that resulted in a minimization of the errors in soiling transmittance estimation. In general, better results were obtained when a three measurement model was employed, compared to a two measurement model.

When used to estimate the soiling losses of different PV materials exposed to the same soiling under different irradiance conditions, the spectral models were found to perform significantly better than single-measurement approaches for most of the PV materials. It was also possible to further reduce the error in the estimation of the soiling ratio for the two-measurement model by optimizing the combinations of wavelengths depending on the PV absorber materials. These models have also been found to be more robust to transmittance measurement errors. In general, the utilization of spectral models is therefore beneficial, especially in high soiling conditions.

From the analysis presented in this paper, one can conclude that optical techniques for monitoring the consequences of soiling in PV could be improved by utilizing two or three wavelengths to estimate the full transmittance spectra, rather than just one. This will make it possible to adjust the soiling measurement, not only to different irradiance conditions but also for different PV module materials. This would enable features currently lacking in both optical sensors and standard soiling stations. This study was conducted at a single location, and should be repeated for a larger number of locations experiencing more significant soiling losses. Future studies should analyse the time series of the transmittance data and soiling loss for each location by using Eqn. 4 and the methods described in this work.

## Acknowledgments

Part of this work was funded through the European Union's Horizon 2020 research and innovation programme under the NoSoilPV project (Marie Skłodowska-Curie grant agreement No. 793120). This work was authored in part by Alliance for Sustainable Energy, LLC, the manager and operator of the National Renewable Energy Laboratory for the U.S. Department of Energy (DOE) under Contract No. DE- AC36-08GO28308. Funding was provided in part by the U.S. Department of Energy's Office of Energy Efficiency and Renewable Energy (EERE) under Solar Energy Technologies Office (SETO) Agreement Number 30311. This study is partially based upon work from COST Action PEARL PV (CA16235), supported by COST (European Cooperation in Science and Technology). COST (European Cooperation in Science and Technology) is a funding agency for research and innovation networks. Our Actions help connect research initiatives across Europe and enable scientists to grow their ideas by sharing them with their peers. This boosts their research, career and innovation, see www.cost.eu.